\documentclass{iopart}
\usepackage{iopams}

\pdfoutput=1

\usepackage{hyperref}
\usepackage{graphicx}
\usepackage{mathptmx}
\usepackage[ngerman,american]{babel}
\usepackage{cite}
\usepackage{url}
\usepackage{longtable}
\usepackage{hhline}
\usepackage{numprint}

\newcommand{\bigo}[1]{O(#1)}

\newcommand{\eqref}[1]{(\ref{#1})}
\newcommand{\text}[1]{\mathrm{#1}}

\newcommand{\fix}{\mathrm{Fix}}

\begin{document}
\title{Low Autocorrelation Binary Sequences}

\author{Tom Packebusch$^1$ and Stephan Mertens$^{1,2}$}

\address{\selectlanguage{ngerman}{$^1$Inst.\ f.\ Theo.\ Physik,
    Otto-von-Guericke Universit"at, PF~4120, 39016 Magdeburg,
    Germany}} 

\address{$^2$Santa Fe Institute,
1399 Hyde Park Rd,
Santa Fe, NM 87501,
USA}

\ead{mertens@ovgu.de}

\begin{abstract}
  Binary sequences with minimal autocorrelations have applications in
  communication engineering, mathematics and computer science. In
  statistical physics they appear as groundstates of the
  Bernasconi model. Finding these sequences is a notoriously hard
  problem, that so far can be solved only by exhaustive search. We
  review recent algorithms and present a new algorithm that finds
  optimal sequences of length $N$ in time $O(N\,1.73^N)$. We
  computed all optimal sequences for $N\leq 66$ and all optimal
  skewsymmetric sequences for $N\leq 119$. 
\end{abstract}

%\pacs{} 

%\submitto{\JPA}

\section{Introduction}

Consider a sequence $S=(s_1,\ldots,s_N)$ with $s_i=\pm 1$. The
autocorrelations of $S$ are defined as  
\begin{equation}
  \label{eq:def-Ck}
  C_k(S) = \sum_{i=1}^{N-k} s_i s_{i+k}
\end{equation}
for $k=0,1,\ldots,N-1$, and the ``energy'' of $S$ is defined as the sum
of the squares of all off-peak correlations,
\begin{equation}
  \label{eq:def-E}
  E(S) = \sum_{k=1}^{N-1} C_k^2(S)\,.
\end{equation}
The \emph{low-autocorrelation binary sequence} (LABS) problem is to
find a sequence $S$ of given length $N$ that minimizes $E(S)$ or, equivalently, 
maximizes the \emph{merit factor}
\begin{equation}
\label{eq:def-merit}
F(S) = \frac{N^2}{2E(S)}\,.
\end{equation}

The LABS problem arises in practical applications in communications 
engineering, where low autocorrelation sequences are used for example as
modulation pulses in radar and sonar ranging
\cite{golay:72,beenker:etal:85,pasha:etal:00}. A particularly exciting application is the
interplanetary radar measurement of spacetime curvature
\cite{shapiro:etal:68}. 

In mathematics, the LABS problem appears in terms of the
Littlewood problem \cite{littlewood:68,borwein:02}, the problem of
constructing polynomials with coefficients $\pm1$  that are ``flat'' on the unit circle in the complex plane.

In statistical physics, $E(S)/N$ can be interpreted as the energy of
$N$ interacting Ising spins $s_i = \pm 1$. This is the Bernasconi
model \cite{bernasconi:87}.  It has long-range 4-spin interactions and
is completely deterministic, i.e. there is no explicit or quenched
disorder like in spin-glasses. Ne\-ver\-the\-less the ground states
are highly disordered -- quasi by definition.  This self-induced
disorder resembles very much the situation in real glasses.  In fact,
the Bernasconi-model exhibits features of a glass transition like a
jump in the specific heat and slow dynamics and
aging \cite{krauth:mezard:95}. A clever variation of the replica
method allows an ana\-ly\-ti\-cal treatment of the Bernasconi model in
the high-temperature regime
\cite{bouchaud:mezard:94,marinari:parisi:ritort:94a}.  For the
low-temperature re\-gime, analytical results are rare -- especially
the ground states are not known. Due to this connection to physics we
refer to the $s_i$ as spins throughout the paper.

These examples illustrate the importance of the LABS
problem in various fields. For more applications and the history of the problem 
we refer to existing surveys \cite{jedwab:survey,hoholdt:06}. In this
contribution we focus on algorithms to solve the LABS problem. But
before we discuss algorithms, we will give a brief survey on what is
known about solutions. 

\section{What is known}

The correlation $C_k$ is the sum of $N-k$ terms $\pm 1$, hence the
value of $|C_k|$ is bounded from below by
\begin{equation}
  \label{eq:bk}
  |C_k| \geq b_k = (N-k) \bmod 2\,.
\end{equation}
A binary sequence with $|C_k| = b_k$ is called a Barker sequence
\cite{barker:53}. The merit factor of a Barker sequence is
\begin{equation}
  \label{eq:merit-barker}
  F_N^{\text{Barker}} = \cases{
    N & for $N$ even, \\
    \frac{N^2}{N-1} & for $N$ odd.
 } 
\end{equation}
If it exists, a Barker sequence is a solution of the LABS
problem. Barker sequences exist for $N=2,3,4,5,7,11$ and $13$, but
probably for no other values of $N$. In fact it can be proven that
there are no Barker sequences for odd values of $N>13$
\cite{turyn:storer:61,schmidt:willms:15}. For even values of $N$, the
existence of Barker sequences can be excluded for $4 < N\leq\numprint{2e30}$
\cite{leung:schmidt:12}.

Let $F_N$ denote the maximum merit factor for sequences of length $N$.
It is an open problem to prove (or disprove) that $F_N$ is bounded.
For Barker sequences, $F_N\propto N$, and the same is true more
generally for sequences such that $|C_k| \leq C^\star$ for some
constant $C^\star$ that does not depend on $N$ or $k$. The common
belief is that no such sequences exist and that $F_N$ is bounded by
some constant.

A non-rigorous argument for $F_N$ being bounded was given by Golay 
\cite{golay:82}. Assuming that the correlations $C_k$ are independent,
he argued that asymptotically $F_N\lesssim 12.3248$, or more precisely, that
\begin{equation}
  \label{eq:golay-bound}
  F_N \lesssim \frac{12.3248}{(8\pi N)^{\frac{3}{2N}}}\,.
\end{equation}

There are some rigorous results for lower bounds on $F_N$.
The mean value of $1/F$, taken over all binary sequences of length
$N$, is $(N-1)/N$ \cite{newmann:byrnes:90}. Hence we expect 
$F_N \geq 1$. In fact one can explicetly construct sequences for all
values of $N$ that have merit factors larger than $1$. The current
record is set by so called appended rotated Legendre sequences with an
asymptotic merit factor of $6.342061\ldots$
\cite{jedwab:katz:schmidt:13,jedwab:katz:schmidt:13a}.

\begin{figure}
  \centering
  \includegraphics[width=\columnwidth]{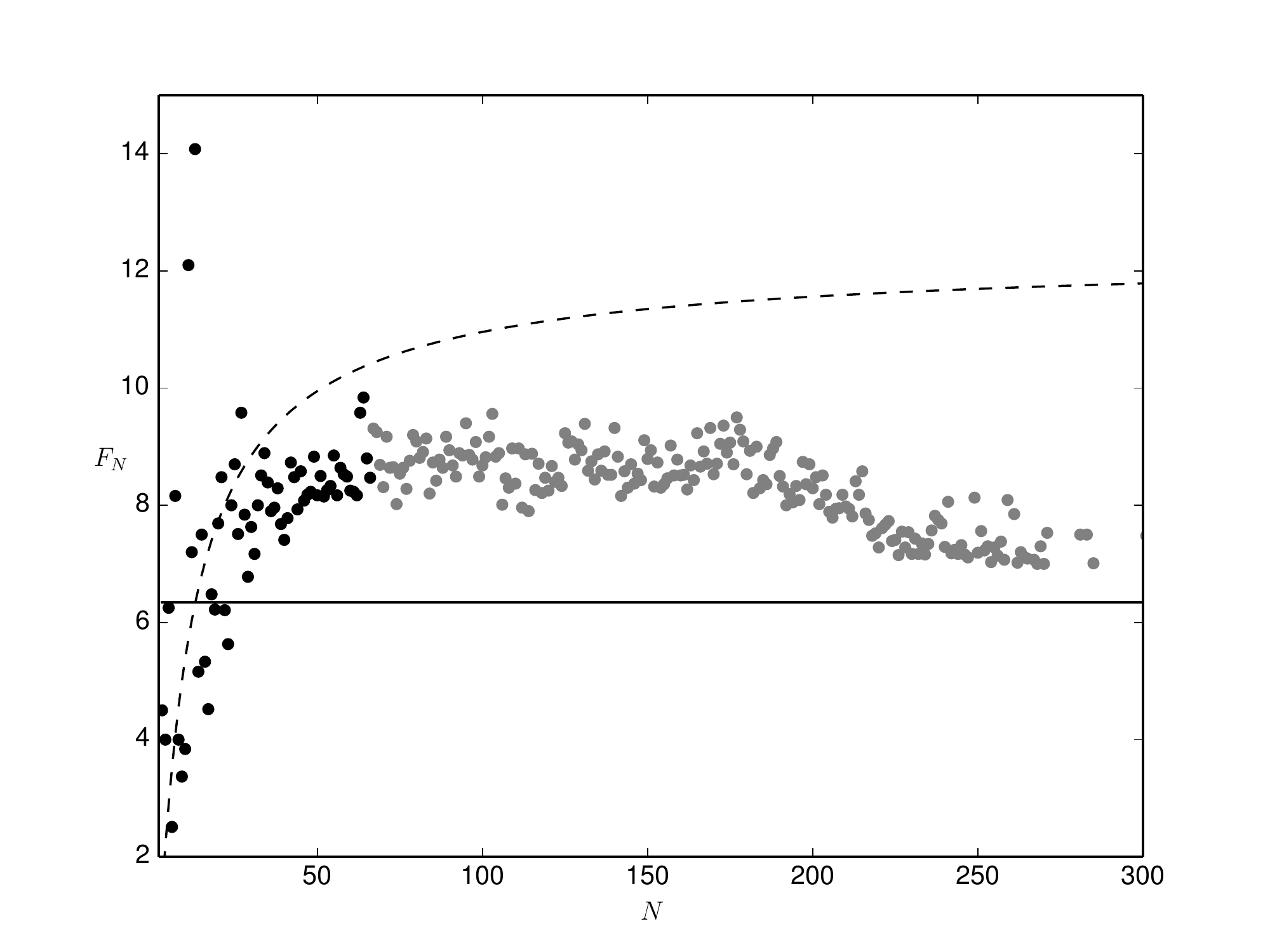}
  \caption{Largest known merit factors. Black symbols are exact
    solutions from exhaustive searches, grey symbols lower bounds from
    heuristic or partial searches. The solid line is the rigorous asymptotic
    lower bound $6.342061\ldots$ from appended rotated Legendre
    sequences \cite{jedwab:katz:schmidt:13a}, the dashed line is
    Golay's non-rigorous asymptotic upper bound
    \eqref{eq:golay-bound}. Data from the tables in
    Section~\ref{sec:results} and from \cite{boskovic:etal:14} and
    references therein.}
  \label{fig:records}
\end{figure}

Beyond that, our knowledge about solutions of the LABS problem is
based on computer searches. Figure~\ref{fig:records} shows the best
merit factors known for $N < 300$. For small
values of $N$, we can exhaustively search through all sequences to
find the sequences with the maximum merit factor $F_N$. 
An evaluation of $E(S)$ from scratch takes time $\Theta(N^2)$, but one
can loop through all sequences sucht that any two successive sequences
differ by exactly one spin, an arrangement known as Gray code \cite{savage:97}. The
corresponding update of $E(S)$ takes only linear time, and the
total time complexity of exhaustive enumeration is then given by $\Theta(N\,2^N)$.
In this paper we will discuss a class of exact enumeration algorithms
with time complexity $\Theta(N\,b^N)$ with $b < 2$ that we used to
solve the LABS problem up to $N\leq 66$.

For larger values of $N$ exhaustive enumeration is not feasible
and one has to resort to either partial enumerations or heuristic
searches. In both cases one obtains sequences with large but not
necessarily maximal merit factors.

Partial enumerations are exhaustive enumerations of a well defined
subset of sequences. A particular promising subset is given by
skewsymmetric sequences of odd length $N=2n-1$. These sequences
satisfy
\begin{equation}
  \label{eq:skew-symmetry}
  s_{n+\ell} = (-1)^\ell s_{n-\ell} \qquad (\ell=1,\ldots,n-1)\,,
\end{equation}
which implies that $C_k=0$ for all odd $k$. The restriction to
skewsymmetric sequences reduces the size of the search space from
$2^N$ to $2^{N/2}$. Sequences with maximum merit factor are often, but
not always
skewsymmetric: from the 31 LABS problems for odd
$N\leq 65$, 21 have skewsymmetric solutions (Section
\ref{sec:results}). For the other values of $N$, skewsymmetric sequences provide lower bounds
for $F_N$. We used our enumeration algorithm to compute the optimal
skewsymmetric sequences for all $N\leq 119$. 
 
Enumerative algorithms (complete or partial) are limited to small
values of $N$ by the exponential size of the search space. Heuristic
algorithms use some plausible rules to locate good sequences more
quickly.  Examples are simulated annealing, evolutionary algorithms,
tabu search---the list of heuristic algorithms that have been applied
to the LABS problem is much longer, see \cite{groot:etal:92}. The
state of the art are the solvers described in \cite{boskovic:etal:14},
which have found many of the merit factors shown in Figure
\ref{fig:records}. The figure shows a significant drop of the
merit factors for $N > 200$. This is generally attributed 
to the fact that even sophisticated search heuristics fail for LABS
problems of larger size. This hardness has earned the LABS problem a place
in CSPLIB, a library of test problems for constraint solvers
\cite[problem 005]{csplib}.

\section{Algorithm}

According to the current state of knowledge, the only way to get exact solutions for the
LABS problem is exhaustive search. With a search space that grows like $2^N$, this
approach is limited to rather small values of $N$, however. The exponential
complexity calls for a method to restrict the search to smaller
subspaces without missing the exact solutions. This is where
branch\&bound comes in, a powerful and versatile method from combinatorial
optimization \cite {noc}. All exact solutions of the LABS problem for $N>32$ 
have been obtained with variations of a 
branch\&bound algorithm proposed in \cite{mertens:96a} that reduces
the size of the search space from $2^N$ to $b^N$ with $b < 2$.
In this section we review these algorithms and we present a new variant 
which has $b=1.72$, the best value to date.

The idea of branch\&bound is to solve
a discrete optimization problem by breaking up its feasible set into successively
smaller subsets ({\em branch}), calculating bounds on the objective
function value over each subset, and using them to discard certain
subsets from further consideration ({\em bound}) \cite{noc}. The procedure ends
when each subset has either produced a feasible solution, or has been
shown to contain no better solution than the one already in hand.  The
best solution found during this procedure is a global optimum.

The goal is of course to discard many subsets as early as possible
during the branching process, i.e.\ to discard most of the feasible
solutions before actually evaluating them. The success of this
approach depends on the branching rule and very much on the quality of
the bound, but it can be quite substantial. 

% With the original branch\&bound
% approach from \cite{mertens:96a}, the scaling of the running time for
% LABS could be reduced from $\bigo{2^N}$ to $\bigo{1.85^N}$. A
% parallelized version of this algorithm was then used used to solve the
% LABS problem for $N\leq 60$ on a 160 CPU Linux cluster
% \cite{bauke:labs}. Later, an improved version was run on 18 GPUs to find
% the maximum merit factors for $N \leq 64$ \cite{wiggenbrock:10}.

For the LABS problem we specify a set of feasible solutions be fixing the $m$ leftmost and
the $m$ rightmost spins of the sequence. The $N-2m$ centre spins
are not specified, i. e.\ the set contains $2^{N-2m}$
feasible solutions. Given a feasible set specified by the $2m$ outer
elements, four smaller sets are created by fixing the elements
$s_{m+1}$ and $s_{N-m}$ to $\pm 1$ and $m$ is increased by $1$. This is applied recursively until
all elements have been fixed. This is the branching rule introduced by
the original branch\&bound algorithm \cite{mertens:96a}, and it is shared by
all later versions. It has the nice property that the long
range correlations are fixed early in the recursion
process. Specifically, if the $m$ left- and rightmost spins are
fixed, all $C_k$ for $k\geq N-m$ are fixed. In addition, this branching
rule supports the computation of lower bounds very well, as we will
see below.

The branching process can be visualized as a tree in which nodes
represent subsets. Each node has four children corresponding to the
four possible ways to set the two spins in the $(m+1)$th shell. The 
branch\&bound algorithm traverses this tree and tries to exclude as
many branches as possible by computing a bound on the energy that can be 
achieved in a branch. The number of nodes actually visited is a
measure of quality for the bound.

\subsection{Bounds}

Bounds are usually obtained by replacing the original problem
over a given subset with an easier (relaxed) problem such that the
solution value of the latter bounds that of the former. A good
relaxation is one that is easy and fast to solve and yields
strong lower bounds. Most often these are conflicting goals.

An obvious relaxation of the LABS problem is given by the problem to
minimize all values $C_k^2$ \emph{independently}. Hence we replace the
original problem
\begin{equation}
\label{eq:original}
E_{\text{min}} = \min_{\text{free}}\left(\sum_{k=1}^{N-1} C_k^2\right)
\end{equation}
by the relaxed version
\begin{equation}
\label{eq:relaxation}
E_{\text{ min}}^* = \sum_{k=1}^{N-1} \min_{\text{free}}(C_k^2) =
\sum_{k=1}^{N-1} \left(\min_{\text{free}}(|C_k|)\right)^2 \leq E_{\text{ min}}\,,
\end{equation}
where ``free'' refers to the $N-2m$ center elements of $s$ that have not
yet been assigned. All previous branch\&bound approaches to LABS considered
$E_{\text{ min}}^*$ to be too expensive to compute and replaced it by a
weaker, but easily computable bound $E_b \leq E_{\text{ min}}^*$
obtained from bounding $\min_{\text{free}} |C_k|$ from below.

\subsubsection{The original bound.}

In the original algorithm \cite{mertens:96a} the bound $E_b$ is
computed by assigning (arbitrary) values to all free spins, thereby
fixing the values for all correlations to $C_k^\star$. Since flipping
a free spin can decrement $|C_k|$ at most by $2$, a lower bound for
$|C_k|$ is given by
\begin{equation}
  \label{eq:C-bound-mertens}
  \min_{\text{free}} |C_k| \geq \max(b_k, |C_k^\star|-2\hat{f}_k)\,,
\end{equation}
where
\begin{equation}
  \label{eq:spin-fk}
  \hat{f}_k =  \cases{
     0 & if $k \geq N-m$,\\
     2(N-m-k) & if $N/2 \leq k < N-m$ or\\
     N-2m & otherwise
  }
\end{equation}
denotes the number of free spins that appear in $C_k$ and $b_k$ is
given by \eqref{eq:bk}. The running time of this algorithm scales like
$\bigo{1.85^N}$. A parallelized version of the algorithm
was used to solve the LABS problem up to $N=60$ \cite{bauke:labs}.

\subsubsection{The Prestwich bound.}

The quality of the bound \eqref{eq:C-bound-mertens} depends on the
values of $C_k^\star$ and hence on the arbitrary values assigned to the
free spins. In principle, these values should be chosen to maximize
$C_k^\star$, but this requires the solution of another optimization
problem for each bound. This can be avoided by considering free
\emph{products} instead of free spins: a product $s_i s_{i+k}$ is free if
$s_i$ or $s_{i+k}$ is a free spin. Products $s_i s_{i+k}$ in which
both spins are fixed ars called fixed. Let $c_k(s)$ denote the sum of
all fixed products that contribute to $C_k$. Note that $c_k = C_k$ for $k \geq N-m$.
Then 
\begin{equation}
  \label{eq:C-bound-prestwich}
  \min_{\text{free}} |C_k| \geq \max(b_k, |c_k(s)|-f_k)\,,
\end{equation}
where
\begin{equation}
  \label{eq:product-fk}
  f_k = (N-k) - 2\max(m-k,0) - \max(k-N+2m,0)
\end{equation}
denotes the number of free products in $C_k$, and
$b_k$ is given by \eqref{eq:bk}.
The reasoning behind \eqref{eq:C-bound-prestwich}
is that the sum $c_k$ of fixed products may be offset by the sum of
free products, which is no greater than $f_k$. If $|c_k(s)| > f_k$
then $|c_k(s)|-f_k$ is a lower bound for $|C_k|$. If
$|c_k(s)| \leq f_k$, this bound is useless and we have to resort to
the trivial lower bound $|C_k(s)|\geq b_k$. The bound
\eqref{eq:C-bound-prestwich} was used by Prestwich to prune parts of the
search space in a local search algorithm for the LABS problem
\cite{prestwich:07}. 

For his recent branch\&{}bound algorithm for LABS, Prestwich
\cite{prestwich:13} improved that bound by taking into account some of
the interactions between fixed and free spins. Suppose that $s_i$ is a
free spin while $s_{i-k}$ and $s_{i+k}$ are fixed. If
$s_{i-k}\neq s_{i+k}$, the contributions
\begin{equation}
  \label{eq:cancellations}
  s_{i-k}s_i + s_i s_{i+k} = s_i (s_{i-k}+s_{i+k})
\end{equation}
of $s_i$ to $C_k$ are zero, no matter what the value of $s_i$ is. For
each such \emph{cancellation}, the number $f_k$ in \eqref{eq:C-bound-prestwich}
can be decreased by two. For $s_{i-k}=s_{i+k}$, the contribution of
the term \eqref{eq:cancellations} is $\pm 2$, a situation referred to as \emph{reinforcement} by
Prestwich. Now, if all free contributions to $C_k$ are either
cancellations or reinforcements, then $f_k$ must be even. If the sum of the fixed
contributions $c_k$ is also even and $c_k\bmod 4 \neq f_k\bmod 4$,
we can set $b_k=2$ in \eqref{eq:C-bound-prestwich}. With this bound,
Prestwich reports a running time that scales like
$\bigo{1.80^N}$. Since Prestwich didn't parallelize his algorithm, this
estimate was based on enumerations only up to $N\leq 44$.

\subsubsection{The Wiggenbrock bound.}

A different bound was used by Wiggenbrock in his
branch\&{}bound algorithm \cite{wiggenbrock:10}. Flipping a spin changes the sum
$C_k+C_{N-k}$ by $\pm 4$ because every spin occurs twice in that sum.
Taking the all $+1$ configuration as a reference, we get
\begin{equation}
  \label{eq:willms}
  (N - C_{N-k}) \equiv C_k \pmod{4}\,.
\end{equation}
For $k \geq N-m$, the $C_k$ are completely fixed. For other values of
$k$, the correlations can be bounded by
\begin{equation}
  \label{eq:C-bound-wiggenbrock}
  |C_k| \geq \cases{
     |(N-C_{N-k}) \bmod 4| & if $k \leq m$, \\
     b_k & if $m < k < N-m$,
  }
\end{equation}
where we assumed the residue system
$\{-1,0,1,2\}$ for the mod 4 operation.

The Wiggenbrock bound seems to be weak since it bounds $|C_k|$ by
small numbers $0,1,2$ only. Yet it is surprisingly efficient:
Wiggenbrock reported a running time of $\bigo{1.79^N}$, slightly
better than the scaling of Prestwich's bound. Using a parallelized
implementation and running it on 18 GPUs, Wiggenbrock solved the 
LABS problem for $N\leq 64$ \cite{wiggenbrock:10}.

\subsubsection{The combined bound.}

High up in the search tree, where $m$ is small, the contributions of
the free products overcompensate the fixed contributions and the
Prestwich bound \eqref{eq:C-bound-prestwich} reduces to $b_k$.
The Wiggenbrock bound \eqref{eq:C-bound-wiggenbrock} provides a better
bound in exactly these situations. The fact that it yields such a good
running time indicates that even this weak bound is efficient because
it applies high up in the search tree: a branch, that can be
pruned at this level, is usually very large.  The Prestwich bound with
the free products applies for larger values of $m$, on the other hand.
An obvious idea is to combine these complimentary bounds and use
\begin{equation}
  \label{eq:C-bound-combined}
  |C_k| \geq \cases{
     \max \big(|(N-C_{N-k}) \bmod 4|, |c_k(s)|-f_k\big) & if $k \leq m$, \\
     \max (b_k, |c_k(s)|-f_k) & if $m < k < N-m$,
  }
\end{equation}
as a bound. 

\begin{figure}
  \centering
  \includegraphics[width=\columnwidth]{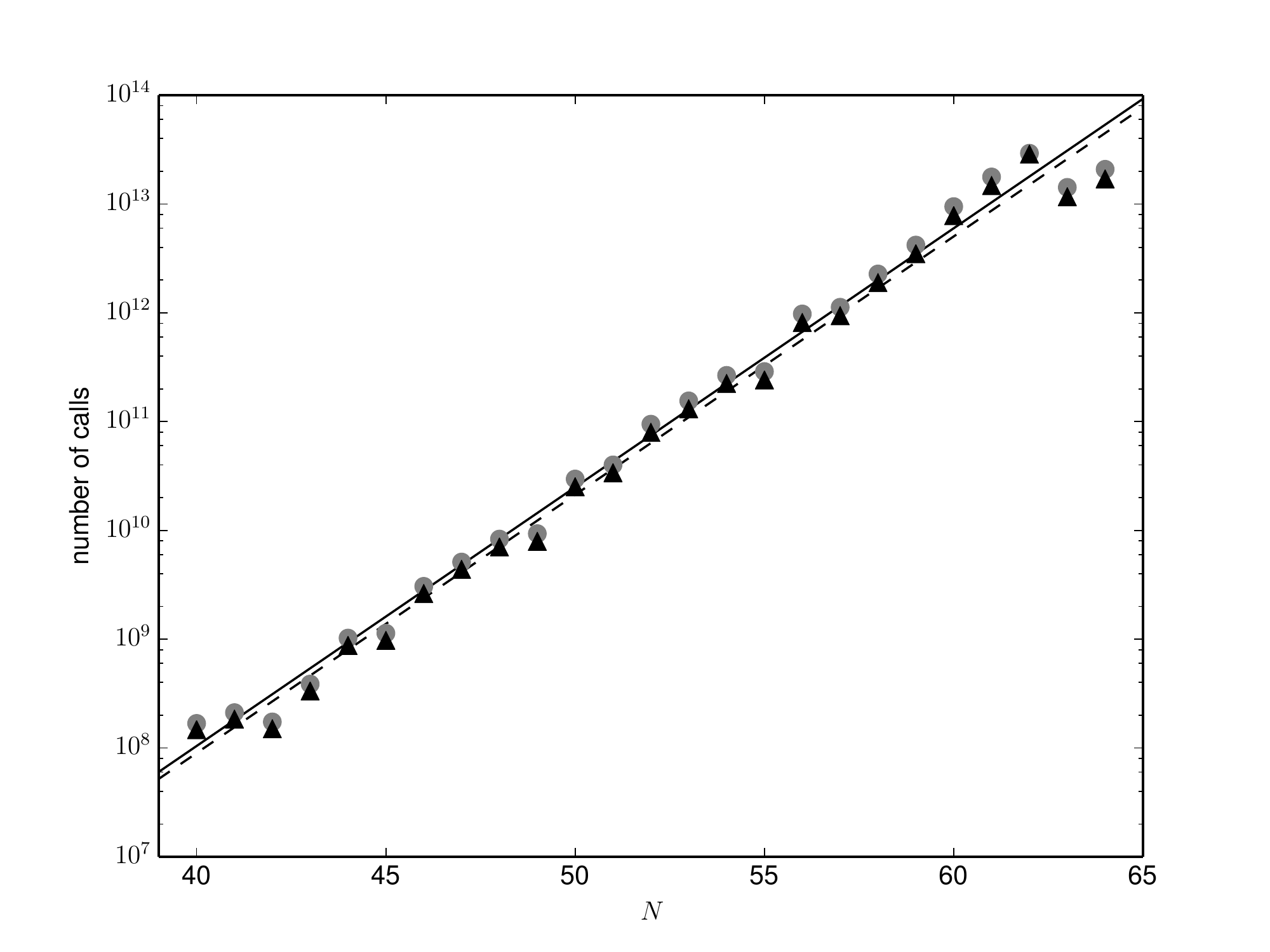}\\
  \includegraphics[width=\columnwidth]{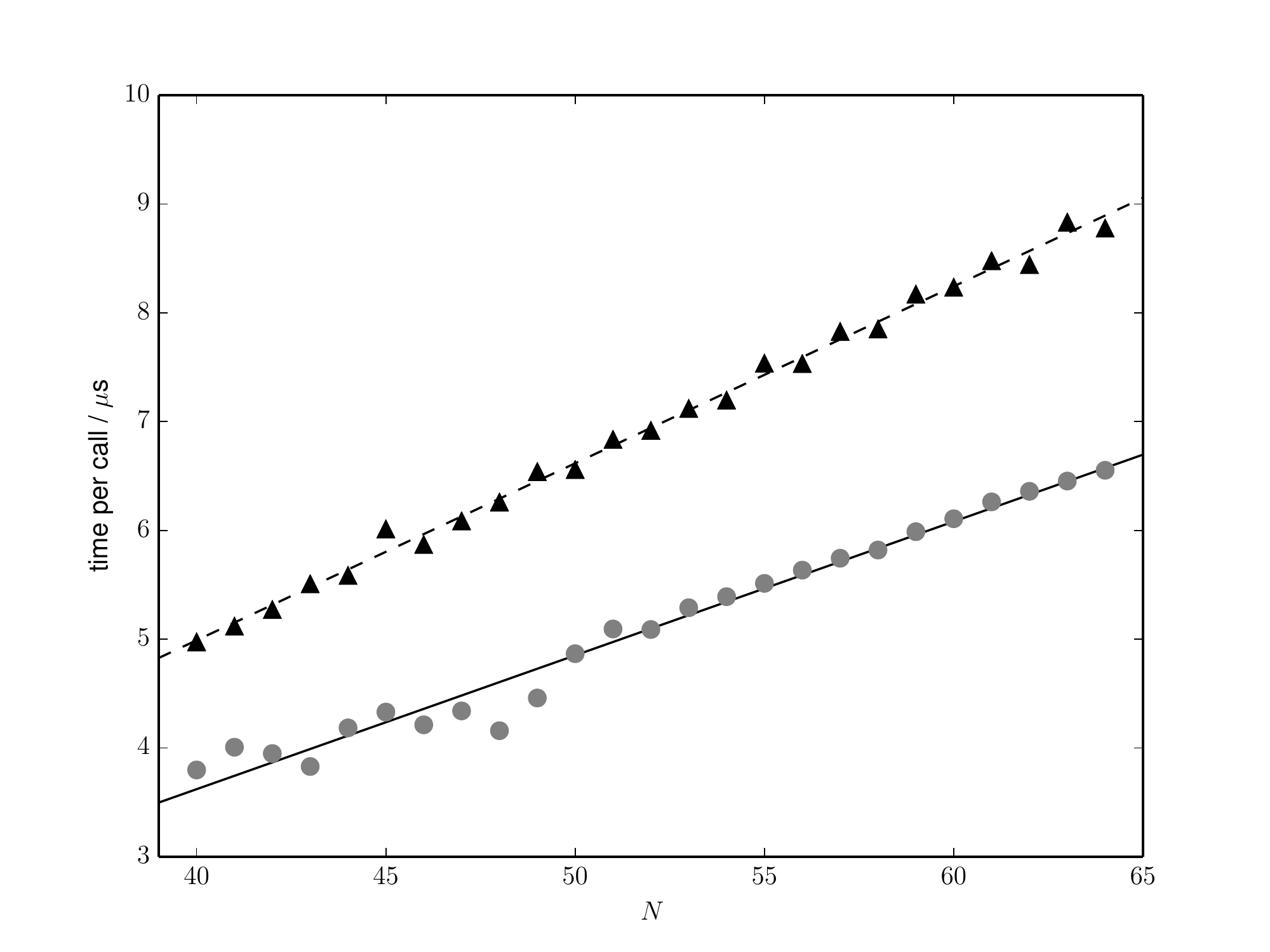}
  \caption{Branch\&{}bound algorithm with the combined bound
    \eqref{eq:C-bound-combined} (circles) and with the tight bound
    \eqref{eq:min-Ck} (triangles). The number of recursive calls (top)
    scales like $\Theta(b^N)$. A numerical fit to the existing data yields $b=1.729$ (solid line) for the
    combined bound and $b=1.727$ (dashed line) for the tight bound,
    but this small difference is caused by a non-exponential reduction of
    the number of calls, see Figure~\ref{fig:ratio}.
    The CPU time per call (bottom) is linear in $N$ for both bounds.
    \label{fig:performance}}
\end{figure}

When we measure the number of recursive calls (i.e. the number of
nodes visited in the search tree) and the CPU time per call
(Figure \ref{fig:performance}), we find that the running
time of the branch\&bound algorithm with the combined bound
\eqref{eq:C-bound-combined} scales like $\Theta(N 1.729^N)$. 

\subsubsection{The tight bound.}

So far, all bounds are lower bounds for $\min|C_k|$ that yield only a lower
bound for $E_{\text{min}}^*$, which already is a lower bound for
$E_{\text{min}}$.  We will now show that $\min|C_k|$ and hence
$E_{\text{min}}^*$ can be computed exactly, thereby avoiding the
``second relaxation'' to $E_b$ and providing the best lower bound possible
from the Ansatz \eqref{eq:relaxation}.  We write $C_k$ as
\begin{equation}
  \label{eq:c-plus-u}
  C_k = c_k + u_k\,,
\end{equation}
where $c_k$ is the sum of all fixed terms $s_i s_{i+k}$ (as above) 
and $u_k$ sums up all terms in which at least one spin is free.
Let 
\begin{equation}
  \label{eq:granularity}
  g_k = \cases{
     4 & if $k \leq m$, \\
     2 & otherwise.
  }
\end{equation}
We will show below that there exist easy to compute integers $U_k^{\text{min}}$ and
$U_k^{\text{max}}$ such that the free contribution $u_k$ can take on all values in 
\begin{equation}
  \label{eq:u-values}
  \{U_k^{\text{min}}, U_k^{\text{min}}+g_k, U_k^{\text{min}}+2g_k, \ldots, U_k^{\text{max}}-g_k, U_k^{\text{max}}\}
\end{equation}
All we need to know are the values of $c_k$, $U_k^{\text{min}}$ and $U_k^{\text{max}}$ to compute 
\begin{equation}
  \label{eq:min-Ck}
  \min |C_k| = \cases{
      c_k + U_k^{\text{min}} & if $-c_k \leq U_k^{\text{min}}$, \\
     c_k+U_k^{\text{max}} & if $-c_k \geq U_k^{\text{max}}$, \\
    |(-c_k-U_k^{\text{min}}) \bmod g_k| & otherwise,
   }
\end{equation}
and then $E_{\text{min}}^* = \sum_k (\min|C_k|)^2$. 
 
To prove \eqref{eq:granularity} and \eqref{eq:u-values}, we rearrange
the sum \eqref{eq:def-Ck} for $C_k$ a little bit. For $C_3$ and $N=12$, for example, we
can write
\begin{eqnarray*}
  C_3 &=& s_1s_4 + s_1s_4 +  s_1s_4 +  s_1s_4 +  s_1s_4 +  s_1s_4 +
        s_1s_4 +  s_1s_4 +  s_1s_4\\
         &=& (s_1s_4 + s_4s_7+s_7s_{10}) + (s_2s_5+s_5s_8+s_8s_{11}) +
    (s_3s_6+s_6s_9+s_9s_{12})\,. 
\end{eqnarray*}
We call every sum in parentheses a \emph{chain}.
For general values of $k$ and $N$ we write
\begin{equation}
  \label{eq:Ck-chains} 
   C_k = \sum_{j=1}^k \sum_{q=1}^{\lfloor\frac{N-j}{k}\rfloor}
  s_{j+(q-1)k} s_{j+qk}\,.
\end{equation}
The chains are the sums over $q$. For
$k<N-m$, each chain contains a subchain of free terms
\begin{equation}
  \label{eq:chain}
  s_a s_{a+k} + s_{a+k}s_{a+2k} + \cdots + s_{b-k} s_b\,
\end{equation}
where only the spins $s_a$ and $s_b$ may be fixed. We refer to these
subchains as free chains. The sum of all free chains equals $u_k$.

Let us first prove the ``granularity'' \eqref{eq:granularity}.
If both spins $s_a$ and $s_b$ are fixed, then every free spin appears
exactly in two terms, and flipping any free spin changes the sum
\eqref{eq:chain} by $0$ or $\pm 4$. If either $s_a$ or $s_b$ (or both) are
free, then flipping this spin changes the sum \eqref{eq:chain} by
$\pm 2$. Hence the granularity $g_k$ is $4$ if and only if all
contributing free chains have both $s_a$ and $s_b$ fixed, and $2$
otherwise.  

Now $s_a$ can be free and the leftmost member of a free
chain if and only if $a>m$ and if it has no left partner, i.e. if
$a-k \leq 0$. Together, both conditions imply $k>m$. Hence by
argumentum e contrario,  $k\leq m$
implies that $s_a$ is fixed and, by similar reasoning, also that
that $s_b$ is fixed. This proves that $g_k = 4$ for $k\leq m$.  

If $k > m$, we only need to find a single free chain that starts with
a free spin. Consider the spin $s_{m+1}$: It is free and it has no
left neighbor. Hence it is the leftmost spin of a free chain that
contributes to $u_k$. Therefore $g_k=2$ for $k > m$. Note that for
$k>m$ there can be free chains with both $s_a$ and $s_b$ fixed. All we
have proven is that for $k > m$ this can't happen for all free chains.

Now we will prove \eqref{eq:u-values}.
Let $n$ denote the number of terms $s_j s_{j+k}$ in a free chain
\eqref{eq:chain}, and let $u$ denote its value. If $s_a$ or $s_b$ (or both) are free, then
$u$ can take on all values between $-n$ and $n$ with
granularity $2$:
\begin{equation}
  \label{eq:u-values-free}
  u \in [-n, -n+2, \ldots, n-2, n]  \qquad \mbox{($s_a$
    or $s_b$ free).}
\end{equation}
If both spins $s_a$ and $s_b$ are fixed, the
granularity is $4$ and the range of values varies with $s_a$, $s_b$ and
the parity of $n$ according to
\begin{equation}
  \label{eq:u-values-fixed}
  u \in \cases{
    [-n,\ldots,n] & if $s_a=s_b$ and $n$ even, \\
    [-(n-2),\ldots,n] & if $s_a=s_b$ and $n$ odd, \\
    [-(n-2),\ldots,(n-2)] & if $s_a\neq s_b$ and $n$ even, \\
    [-n,\ldots,(n-2)] & if $s_a\neq s_b$ and $n$ odd.
  }
\end{equation}
This can be proven by induction over $n$. For $n$ odd, the base
case is $n=3$, i.e.
\begin{displaymath}
  u = s_a s_{a+k} + s_{a+k} s_{b-k} + s_{b-k}s_b\,.
\end{displaymath}
The value of $u$ is maximized by setting the free spins $s_{a+k}=s_a$
and $s_{b-k}=s_b$. If $s_a=s_b$, the center term is $1$ and $u_{\text{max}}=3$. For
$s_a\neq s_b$, the center term is $-1$ and $u_{\text{max}}=1$. The value of $u$ is
minimized by setting $s_{a+k}=-s_a$
and $s_{b-k}=-s_b$. If $s_a=s_b$, the center term is $1$ and
$u_{\text{min}}=-1$. If $s_a\neq s_b$, the center term is $-1$ and $u_{\text{min}}=-3$.
Now let us assume that \eqref{eq:u-values-fixed} holds for some odd $n
\geq 3$ and consider a free chain 
\begin{displaymath}
  u = s_a s_{a+k} + s_{a+k} s_{a+2k} + \cdots + s_{b-2k}s_{b-k} + s_{b-k}s_b
\end{displaymath}
with $n+2$ terms.
To maximize $u$, we set $s_{a+k}=s_a$ and $s_{b-k}=s_b$, and the
remaining free chain has $n$ terms. Applying
\eqref{eq:u-values-fixed}, we get $u_{\text{max}} = n+2$ if $s_a=s_b$
and $u_{\text{max}} = n$ if $s_a\neq s_b$. The induction step for
$u_{\text{min}}$ is obvious.

Since the proof for even $n$ is very similar, it is omitted here.
We only mention that the base case ($n=2$) corresponds to the
``cancellation'' and ``reinforcement'' used by Prestwich to improve the
bound \eqref{eq:C-bound-prestwich}.

Now \eqref{eq:u-values-free} and \eqref{eq:u-values-fixed} tell us how
to compute $u_{\text{min}}$ and $u_{\text{max}}$ for each individual
free chain. The corresponding values $U_k^{\text{min}}$ and
$U_k^{\text{max}}$ are obtained by summing over all free chains that
contribute to $u_k$.

\begin{figure}
  \centering
  \includegraphics[width=\columnwidth]{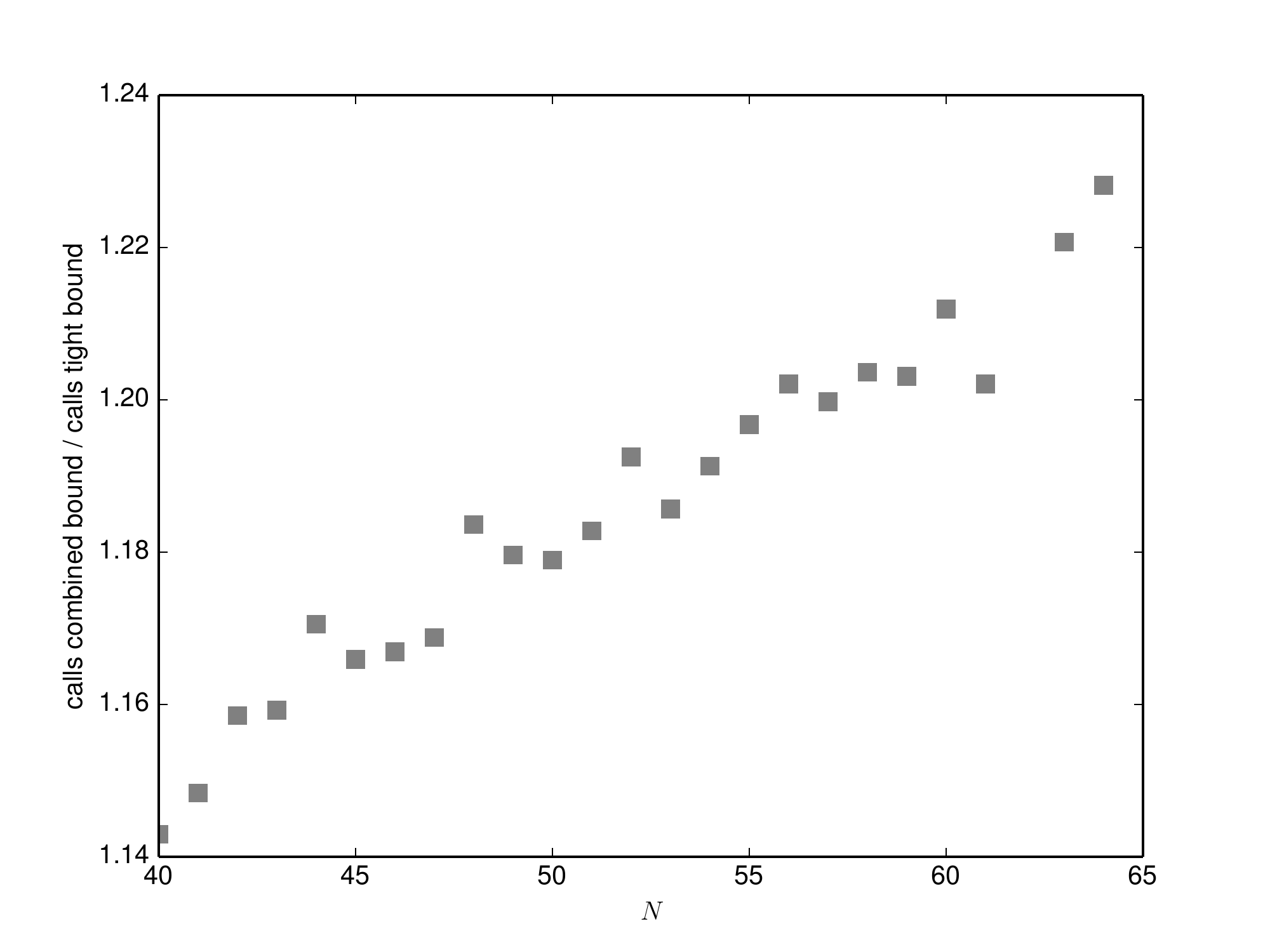}\\
  \caption{Number of calls for the combined bound
    \eqref{eq:C-bound-combined}  divided by the number of calls for the tight bound
    \eqref{eq:min-Ck}. For the values of $N$ considered here, the
    speedup due to the tight bound seems to grow linearly with $N$.
    \label{fig:ratio}}
\end{figure}

Every branch of the search tree that can be pruned according to the
combined bound \eqref{eq:C-bound-combined} (or any other relaxation of
\eqref{eq:relaxation}) is also pruned by the tight bound
\eqref{eq:min-Ck}, but the tight bound allows us to prune additional
branches. Hence the number of recursive calls with the tight bound
can not be larger than the number of calls with any other bound based
on \eqref{eq:relaxation}.  What we observe is that for $N\leq 66$ the number of calls for
the tight bound is in fact strictly smaller than that for the combined
bound.  A numerical fit to the
existing data yields a scaling of $\Theta(1.727^N)$ for the tight
bound, compared to $\Theta(1.729^N)$ for the combined bound, see
Figure \ref{fig:performance}. This difference is too small to tell whether 
the tight bound actually provides an exponential speedup or not.  In fact, if one looks at the
ratio of the number of calls for the combined bound divided by the
number of calls for the tight bound, one observes that the speedup factor
grows linearly with $N$, not exponentially (Figure~\ref{fig:ratio}). 
Since the time per call scales linearly for both bounds
(Figure~\ref{fig:performance} bottom), a reduction of the number of
calls that grows with $N$ implies that the
tight bound will asymptotically outperform the combined bound.

For the values of $N$ considered in this paper, however, the absolute computational costs
per call matter. And here the simpler combined bound
\eqref{eq:C-bound-combined} is faster, see
Figure~\ref{fig:performance} (bottom). If we extrapolate the number of calls and
the time per call to $N=66$, we get a running time of roughly 12600
CPU days for the combined bound but 14300 CPU days for the tight
bound. This is why we used the weaker combined bound for all the new
solutions (exact and skewsymmetric) reported in this paper. Note that
the time per call depends considerably on the implementation. It might
well be possible to implement the tight bound such that it outperforms
the combined bound already for the values of $N$ considered here. In
any case, the measured running times illustrate that we need to
parallelize the computation if we don't want to wait 35 years for the
$N=66$ LABS solution.

\subsection{Symmetry and Parallelization}

The correlations $C_k$ \eqref{eq:def-Ck} are unchanged when the
sequence is complemented or reversed. When alternate elements of the
sequence are complemented, the even-indexed correlations are not
affected, the odd-indexed correlations only change sign. Hence, with
the exception of a small number of symmetric sequences, the $2^N$
sequences will come in classes of eight which are equivalent. The
total number of nonequivalent sequences is slightly larger than
$2^{N-3}$.

The $m$ left- and $m$ rightmost elements of the sequence can be used
to parameterize the symmetry classes. The total number $c(m)$ of
symmetry classes that can be
distinguished by $m$ left- and $m$ right-border elements reads
\begin{equation}
   \label{eq:m-classes}
   c(m) = 2^{2m-3} + 2^{m-2+(N\bmod2)}\,.
\end{equation}
We derive this formula in \ref{sec:group-theory}, where we also
describe how to compute the values of the $2m$ boundary spins that
represent each symmetry class. 

The symmetry classes can be enumerated independently, which allows us
to parallelize the computation.  For our largest system ($N=66$) we used $c(m=10)=\numprint{131328}$
symmetry classes that we searched in parallel on various computers
with number of computing cores ranging from \numprint{8} to
\numprint{5700}. In principle, the branch\&bound algorithm requires
some communication between the parallel tasks since every task should
know the lowest energy found so far by other tasks to compare it to the bound. We
avoid this communication completely by using a static value for this reference
energy: the lowest energy found by heuristic searches. In all cases we
considered, this value turned out to be the true minimum energy.
  
\section{Results and Conclusions}
\label{sec:results}

We have used the branch\&bound algorithm with the combined bound to
compute all sequences with maximum merit factor for $N\leq 66$, see
Tables \ref{tab:labs-1} and \ref{tab:labs-2}.  The previous record was
$N\leq 64$, obtained with the Wiggenbrock bound
\eqref{eq:C-bound-wiggenbrock} and using 18 GPUs \cite{wiggenbrock:10}.
For the perfomance measurements for $40\leq N \leq 64$ shown in Figure~\ref{fig:performance}
we have used a Linux cluster with a collection of
Intel\textsuperscript{\textregistered}\
Xeon\textsuperscript{\textregistered}\ CPUs: $10\times$ E5-2630 (at
2.30 Ghz), $10 \times$ E5-2630 v2 (at 2.60 Ghz) and
$2\times$ E5-1620 (at 3.60 Ghz) with a
total of 248 (virtual) cores. On this machine, the computation for
$N=64$ took about a week (wallclock time). As one can see in
Figure~\ref{fig:performance}, the solution of $N=63$ and $N=64$
involves a surprisingly low number of calls and took therefore less
time than actually expected.

Note that with our algorithm systems of size $N\leq 43$ can be solved in less than an hour on a laptop.

For $N=65$ and $N=66$ we used a variety of computing machinery that
makes an accurate determination of ``single CPU time'' impossible. For
$N=65$ and 66, the equivalent wallclock time on our benchmark cluster
is roughly 32 and 55 days.   

\begin{table}
  \parbox{0.23\linewidth}{
  \centering
  \small
\begin{tabular}{rrrlc}
  $N$ & $E$ & $F_N$ & sequences & skew \\[1ex]
3 & 1 & 4.500 & 21 & $\times$ \\
4 & 2 & 4.000 & 112 &   \\
5 & 2 & 6.250 & 311 & $\times$ \\
6 & 7 & 2.571 & 141 &   \\
& & & 123 &   \\
& & & 312 &   \\
& & & 1113 &   \\
7 & 3 & 8.167 & 1123 & $\times$ \\
8 & 8 & 4.000 & 32111 &   \\
& & & 31121 &   \\
9 & 12 & 3.375 & 311121 &   \\
& & & 42111 & $\times$ \\
& & & 32211 & $\times$ \\
& & & 31122 &   \\
10 & 13 & 3.846 & 42211 &   \\
& & & 52111 &   \\
& & & 311122 &   \\
& & & 41122 &   \\
& & & 33121 &   \\
11 & 5 & 12.100 & 112133 & $\times$ \\
12 & 10 & 7.200 & 4221111 &   \\
& & & 4111221 &   \\
13 & 6 & 14.083 & 5221111 & $\times$ \\
14 & 19 & 5.158 & 41112221 &   \\
& & & 6221111 &   \\
& & & 5222111 &   \\
& & & 33111212 &   \\
& & & 41111222 &   \\
& & & 42211112 &   \\
& & & 5221112 &   \\
& & & 5311121 &   \\
15 & 15 & 7.500 & 52221111 & $\times$ \\
& & & 33131211 & $\times$ \\
16 & 24 & 5.333 & 225111121 &   \\
& & & 6322111 &   \\
& & & 313311211 &   \\
& & & 2131441 &   \\
17 & 32 & 4.516 & 252211121 & $\times$ \\
& & & 44121311 &   \\
& & & 4221211112 &   \\
& & & 36111221 &   \\
& & & 2122411112 &   \\
& & & 2112113132 &   \\
18 & 25 & 6.480 & 441112221 &   \\
& & & 511211322 &   
\end{tabular} 
  }
  \hfill
  \parbox{0.47\linewidth}{
  \centering
  \small
\begin{tabular}{rrrlc}
  $N$ & $E$ & $F_N$ & sequences & skew \\[1ex]
19 & 29 & 6.224 & 4111142212 &   \\
20 & 26 & 7.692 & 5113112321 &   \\
21 & 26 & 8.481 & 27221111121 & $\times$ \\
22 & 39 & 6.205 & 51221111233 &   \\
& & & 632111112211 &   \\
& & & 511111212232 &   \\
23 & 47 & 5.628 & 212121111632 &   \\
& & & 83211112211 &   \\
& & & 314121131132 &   \\
24 & 36 & 8.000 & 2236111112121 &   \\
25 & 36 & 8.681 & 337111121221 &   \\
26 & 45 & 7.511 & 21212111116322 &   \\
& & & 63231111121211 &   \\
& & & 32361111121211 &   \\
27 & 37 & 9.851 & 34313131211211 & $\times$ \\
28 & 50 & 7.840 & 34313131211212 &   \\
29 & 62 & 6.782 & 212112131313431 & $\times$ \\
& & & 323711111212211 & $\times$ \\
30 & 59 & 7.627 & 551212111113231 &   \\
& & & 461212111113231 &   \\
31 & 67 & 7.172 & 7332212211112111 &   \\
32 & 64 & 8.000 & 71112111133221221 &   \\
33 & 64 & 8.508 & 742112111111122221 &   \\
34 & 65 & 8.892 & 842112111111122221 &   \\
35 & 73 & 8.390 & 7122122111121111332 &   \\
36 & 82 & 7.902 & 3632311131212111211 &   \\
37 & 86 & 7.959 & 844211211111122221 &   \\
38 & 87 & 8.299 & 8442112111111122221 &   \\
39 & 99 & 7.682 & 82121121234321111111 & $\times$ \\
& & & 23241171111141122121 & $\times$ \\
40 & 108 & 7.407 & 44412112131121313131 &   \\
41 & 108 & 7.782 & 343111111222281211211 & $\times$ \\
42 & 101 & 8.733 & 313131341343112112112 &   \\
43 & 109 & 8.482 & 1132432111117212112213 & $\times$ \\
44 & 122 & 7.934 & 525313113111222111211121 &   \\
45 & 118 & 8.581 & 82121121231234321111111 & $\times$ \\
46 & 131 & 8.076 & 823431231211212211111111 &   \\
& & & 821211212312343211111111 &   \\
& & & 73235111112132122112121 &   \\
47 & 135 & 8.181 & 923431231211212211111111 & $\times$ \\
& & & 429422222112111111122111 & $\times$ \\
& & & 411121114131131312421242 &   \\
& & & 383422132211212111111211 & $\times$ \\
& & & 236331611113121211112121 & $\times$ \\
& & & 212a21121234211111111231 & $\times$ 
\end{tabular}
}
  \caption{All optimal low autocorrelation binary sequences for $N\leq
    47$ modulo symmetries.}
  \label{tab:labs-1}
\end{table}

\begin{table}
  \centering
  {\small
\begin{tabular}{rrrlc}
  $N$ & $E$ & $F_N$ & sequences & skew \\[1ex]
48 & 140 & 8.229 & 3111111832143212221121121 &   \\
49 & 136 & 8.827 & 215131311224112241141141 &   \\
& & & 3337313221312111112121211 & $\times$ \\
50 & 153 & 8.170 & 215131311224112241141142 &   \\
& & & 72542221311111132111211211 &   \\
& & & 4337313221312111112121211 &   \\
51 & 153 & 8.500 & 23432111141313116212112121 & $\times$ \\
52 & 166 & 8.145 & 51161212121111131223123332 &   \\
53 & 170 & 8.262 & 4511311133251312221112111121 &   \\
& & & 22b442222112112111111111221 & $\times$ \\
54 & 175 & 8.331 & 356225141212112222111111121 &   \\
55 & 171 & 8.845 & 9212123212114321233211111111 & $\times$ \\
& & & 3232a41124112111111112212211 & $\times$ \\
56 & 192 & 8.167 & 7612231123241111132112122111 &   \\
57 & 188 & 8.641 & 33232631111127121111221221211 & $\times$ \\
58 & 197 & 8.538 & 1111131232138142121132432112 &   \\
59 & 205 & 8.490 & 772412242112231122111112111111 & $\times$ \\
& & & 6132123121111113112341221121242 &   \\
60 & 218 & 8.257 & 761112141111131124211322211222 &   \\
& & & 222222111311114244161121161121 &   \\
61 & 226 & 8.232 & 314162331211111131112125621211 &   \\
62 & 235 & 8.179 & 323232111117111541121511222122 &   \\
& & & a23223212135311221111112113112 &   \\
63 & 207 & 9.587 & 212212212711111511121143111422321 &   \\
64 & 208 & 9.846 & 212212212711111511121143111422322 &   \\
65 & 240 & 8.802 & 323224111341121115111117212212212 &   \\
66 & 257 & 8.475 & 2112111211222b2221111111112224542 &   
\end{tabular}
  }
  \caption{All optimal low autocorrelation binary sequences for $48
    \leq N\leq 66$ modulo symmetries.}
  \label{tab:labs-2}
\end{table}

Tables~\ref{tab:labs-1} and \ref{tab:labs-2} show all sequences
(except those related by symmetries) with maximum merit factors up to $N=66$ in
run-length encoding, i.e. the digits specify the length of runs of
equal spins. We use $a=10$, $b=11$ etc. for runs of spins that are
longer than $9$.

We have used our branch\&bound algorithm also to find all
skewsymmetric sequences with maximum merit factor up to $N=119$. The previous record was $N\leq
89$ \cite{prestwich:13}. Table \ref{tab:skew} shows the
skewsymmetric sequences with maximum merit factor as far as they are not
listed in Tables \ref{tab:labs-1} and \ref{tab:labs-2}. 
Skewsymmetric merit factors marked with 
$\star$ are known to be not maximal. We know this either from
exhaustive enumerations (for $N\leq 65$) or from heuristic searches
that have yielded non skewsymmetric sequences with larger merit factors.

\begin{figure}
  \centering
  \includegraphics[width=\linewidth]{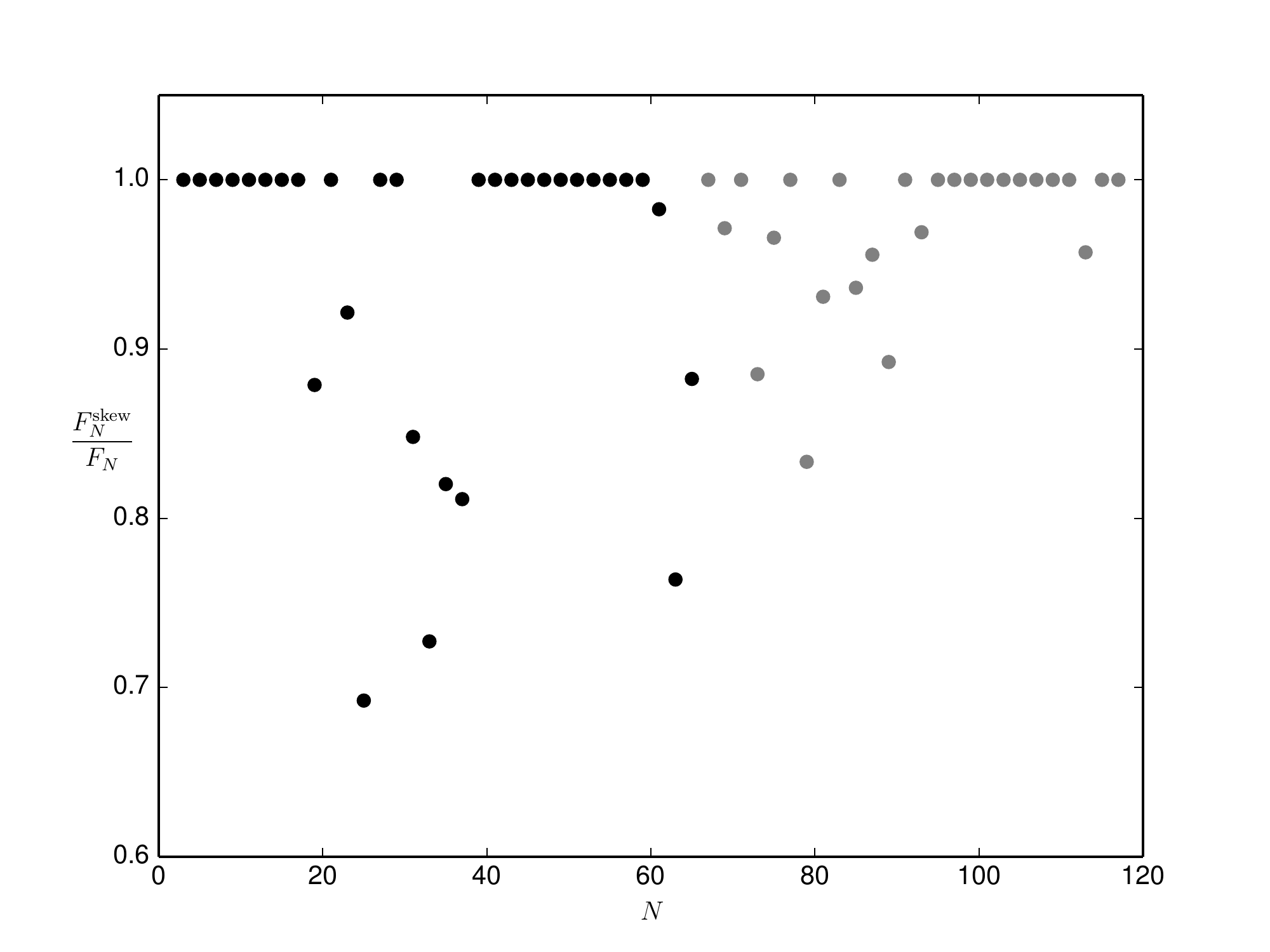}
  \caption{Ratio of maximum merit factors: skewsymmetric
    $F_N^{\text{skew}}$ versus general $F_N$. Black symbols are exact,
    gray symbols are based on lower bounds for $F_N$, which are
    believed to be exact.}
  \label{fig:F-ratio}
\end{figure}

Figure~\ref{fig:F-ratio} shows the ratio of the maximum merit factors
of skewsymmetric and general sequences for $N\leq 119$. In 20 out of
58 cases the skewsymmetric subset does not contain a maximum merit
factor sequence. Note that the values of $F_N$ for $N>66$ are from
heuristic searches, but we believe that these values are the true maximum merit
factors. But strictly speaking, the gray symbols in
Figure~\ref{fig:F-ratio} are only upper bounds for the ratio
$F_N^{\text{skew}}/F_N$.  The available data seems to indicate that
roughly two thirds of all odd values of $N$ have skewsymmetric maximum
merit factor sequences. Figure~\ref{fig:F-ratio} also suggests that
\begin{equation}
  \label{eq:F-ration}
  \liminf_{N\to\infty}  \frac{F_N^{\text{skew}}}{F_N} = 1\,.
\end{equation}

We think that the branch\&bound approach based on the
relaxation \eqref{eq:relaxation} can be used to solve the LABS problem for $N>66$
by devoting more compute cores and more CPU time. Improving the
implementation to reduce the constant factor in the $\Theta(N\,b^N)$
scaling can also help. Solving systems significantly larger than
$N=66$, however, requires a stronger bound than \eqref{eq:relaxation},
i.e., a bound that takes into account the fact that the $C_k$ are not
independent. Or a completely new approach other than branch\&bound.

\begin{table}
  \centering
  {\small
\begin{tabular}{rrrl}
  $N$ & $E$ & $F_N^{\text{skew}}$ & sequences \\[1ex]
19 & 33 & 5.470$^\star$ & 2113114141, 3513111211 \\
23 & 51 & 5.186$^\star$ & 272221111121, 336111121211, 343131211211, 732212111111 \\
25 & 52 & 6.010$^\star$ & 6332121211111 \\
31 & 79 & 6.082$^\star$ & 6212211423211111 \\
33 & 88 & 6.188$^\star$ & 84212321121111111, 22742211211111221 \\
35 & 89 & 6.882$^\star$ & 472322122111112111, 552212232211121111 \\
37 & 106 & 6.458$^\star$ & 2492222111111121121 \\
61 & 230 & 8.089$^\star$ & 2121111221121411111122811342631 \\
63 & 271 & 7.323$^\star$ & a1121112112221222322454111111111 \\
& & & 21242131111311112112461613211231 \\
& & & 23111131111323531211121221616121 \\
65 & 272 & 7.767$^\star$ & 414411126121313133111125112113111 \\
& & & 231134321111114222211821211214121 \\
& & & 221111121211111132311224122183721 \\
& & & 2112111211222b2221111111112224541 \\
67 & 241 & 9.313\phantom{$^\star$} & b412323441121121221231121111111111 \\
& & & 6216121225331212111223311113211111 \\
& & & 2454222111111111222b22211211121121 \\
69 & 282 & 8.441$^\star$  & 211111111121132122121121144323214b1 \\
71 & 275 & 9.165\phantom{$^\star$} & 241244124172222111113112311211231121 \\
73 & 348 & 7.657$^\star$  & 2111211211111221131113213132151427451 \\
& & & 22c7442222221121121111121111111111221 \\
75 & 341 & 8.248$^\star$  & 23231233481611113111111211212123122121 \\
77 & 358 & 8.281\phantom{$^\star$} & 512174112122112221322423411211111331111 \\
79 & 407 & 7.667$^\star$  & 4361113231311213321213413122151111212111 \\
& & & 3121312121411112131112112451361133313311 \\
& & & 2131211221311121211121131141453513243131 \\
& & & 2129214121112121311241335311321111111231 \\
& & & 2111213111121123314261111221131212461351 \\
81 & 400 & 8.201$^\star$  & 53611132313112133212134131221511112121111 \\
83 & 377 & 9.137\phantom{$^\star$} & 323633231172611112211111412212121111212211 \\
85 & 442 & 8.173$^\star$  & 3912523121213351112121333122111231111111211 \\
87 & 451 & 8.391$^\star$  & 43114242215111132131313216111322112211412111 \\
89 & 484 & 8.183$^\star$  & 231143113311111143233221212212118121412114121 \\
& & & 231433161111121421112123521137111131212113121 \\
91 & 477 & 8.680\phantom{$^\star$} & 2121416112211211111211321222321474241111311331 \\
93 & 502 & 8.615$^\star$  & 91252112312341122322122411212312421112311111111 \\
& & & 21121213121261171226221111223111114111123313341 \\
& & & 25523581113122413112231511111121112122111211121 \\
95 & 479 & 9.421\phantom{$^\star$} & 322322358115111351112151114111111211121222122211 \\
97 & 536 & 8.777\phantom{$^\star$} & 5111415321132221132143121132142221421211131151111 \\
99 & 577 & 8.493\phantom{$^\star$} & 5255212212a311224112241211111111232321112111221111 \\
101 & 578 & 8.824\phantom{$^\star$} & 6255212212a3112241122412111111112323211121112211111 \\
103 & 555 & 9.558\phantom{$^\star$} & 2452681222213111225111225132223111111211112211121121 \\
105 & 620 & 8.891\phantom{$^\star$} & a1211121121411213112132221223222134134113453111111111 \\
107 & 677 & 8.456\phantom{$^\star$} & 227311831111224113342221121214112261211111141211111221 \\
109 & 662 & 8.974\phantom{$^\star$} & 3341111112431141111133222251112222212171141211281121211 \\
111 & 687 & 8.967\phantom{$^\star$} & 21111323331321111135114211332121421141112172131212122161 \\
113 & 752 & 8.490$^\star$  & 4555122142121212c1222311111111112333211323111211121112111 \\
& & & 231332171323311541212112134331121114121221311111321213121 \\
115 & 745 & 8.876\phantom{$^\star$} & 5511135145311122113121222222233142512111211311121511121111 \\
117 & 786 & 8.708\phantom{$^\star$} &
                                      37117312111221111133222424112211222212172531211111411111211
  \\
119 & 835 & 8.480\phantom{$^\star$} & 312161412122123411121111314111321511316511212323311311113311
\end{tabular}
  }
  \caption{All optimal skewsymmetric low autocorrelation binary
    sequences for $N\leq 119$ as far as they are not listed in Table
    \ref{tab:labs-1} or \ref{tab:labs-2}. Merit factors marked with 
  $\star$ are known to be not maximal, either from exhaustive
  enumeration (for $N\leq 65$) or from heuristic searches (for $N\geq 67$). }
  \label{tab:skew}
\end{table}

\appendix
\section{Symmetry}
\label{sec:group-theory}

To find the exact number of symmetry classes in the LABS problem we need some group theory.
The operators $R$ (reverse), $C$ (complement) and $A$ (alternate complement) act on the sequences,
leaving the energy invariant. Together with the identity operator
$I$ these operators generate a group $G$ of order $8$. The structure of $G$ depends on
$N$ being odd or even.

For $N$ even, $G=G_e$ is non-abelian and
isomorphic to the dihedral group $D_4$, the symmetry group of a square
plate, generated by a 90 degree rotation and a flip.
The elements of $G_e$ are $\{I,R,C,A,RA,AR,RC,AC\}$. The group elements act on 
a sequence $s$. Let
\begin{equation}
  \label{eq:def-orbit}
  G(s) := \{g(s) : g\in G\}
\end{equation}
denote the orbit of $s$, i.e.\ the set of all spin sequences that can be generated from
$s$ by application of the group elements. The length of the orbit is $|G(s)|$. The orbits
partition the set of all spin sequences in the symmetry classes we want to count. If all
orbits were of length $8$, we would have $2^{N-3}$ symmetry classes. Unfortunately
there are orbits of smaller length, to wit
\begin{displaymath}
  G(++++) = \{++++, ----, -+-+, +-+-\}.
\end{displaymath}
The true number $c$ of orbits is given by Burnside's Lemma,
\begin{equation}
  \label{eq:burnside}
  c = \frac{1}{|G|} \sum_{g\in G} |\fix(g)|\,,
\end{equation}
where $\fix(g) = \{s : g(s)=s\}$ denotes the set of all sequences $s$
that are fixed points of $g$.
The group elements $A$, $C$, $AR$, $RA$ and $AC$ can't fix a sequence, but $R$
and $RC$ can.  Sequences with $s_j = s_{N+1-j}$ 
are fixed by $R$, sequences with $s_j = -s_{N+1-j}$ 
are fixed by $RC$, and there are $2^{N/2}$ sequences of each type. Hence 
\begin{equation}
  \label{eq:sym-classes-even-N}
  c_{N \mathrm{even}} = \frac{1}{8}\left(|\fix(I)| + |\fix(R)| + |\fix(RC)|\right) = 2^{N-3} + 2^{N/2-2}.
\end{equation}
For $N$ odd, the group $G=G_o$ is again of order $8$, but this time it is abelian.
Group elements are $G_o=\{I,R,C,A,RC,RA,CA,RCA\}$, and $g^2=I$ for all $g\in G_o$.
$G_o$ is isomorphic to the reflection-symmetry group of the cube. If
$(N-1)/2$ is odd, only
$R$, $RA$ and $I$ have fixed points. Sequences that are fixed by $R$ have $s_j =
s_{N+1-j}$ with arbitrary center spin $s_{(N+1)/2}$. There are $2\cdot2^{(N-1)/2}$ such sequences.
The same number of sequences are fixed by $RA$. Hence 
\begin{equation}
  \label{eq:sym-classes-odd-N}
  c_{N \mathrm{odd}} = \frac{1}{8}\left( |\fix(I)| + |\fix(R)| + |\fix(RA)| \right) = 2^{N-3} + 2^{(N-1)/2-1}.
\end{equation}
If $(N-1)/2$ is even, only $I$, $R$ and $RCA$ have fixed points, and
their numbers are the same as in \eqref{eq:sym-classes-odd-N}.
Combining \eqref{eq:sym-classes-even-N} and
\eqref{eq:sym-classes-odd-N} provides us with
\begin{equation}
  \label{eq:sym-classes-N}
  c_N = 2^{N-3} + 2^{(N-1)/2-2 + (N \bmod 2)}\,. 
\end{equation}
This is the total number of symmetry classes if we consider all
elements of the sequence. If we only consider the $m$ leftmost and $m$
rightmost elements, the arguments are similar. For $N$ even, 
the symmetry group $G_e=\{I,R,C,A,RA,AR,RC,AC\}$ acts only on the $2m$
elements, and only $I$, $R$ and $RC$ have fixed points. Hence
\begin{equation}
  \label{eq:m-classes-even-N}
  c(m) = 2^{2m-3} + 2^{m-2} \qquad\mbox{$N$ even.}
\end{equation}
For $N$ odd, the symmetry group is again $G_o=\{I,R,C,A,RC,RA,CA,RCA\}$,
but this time $I$, $R$, $RC$, $RA$ and $RCA$ have fixed points:
\begin{equation}
  \label{eq:m-classes-odd-N}
  c(m) = 2^{2m-3} + 2^{m-1} \qquad\mbox{$N$ odd.}
\end{equation}
Combining \eqref{eq:m-classes-even-N} and \eqref{eq:m-classes-odd-N}
provides us with \eqref{eq:m-classes}.

The $c(m)$ symmetry classes can be uniquely parameterized by
the values of the $2m$ boundary spins. Consider the list of all
$2^{2m}$ possible configurations of the boundary spins. For each such
configuration compute $G(s)=G_e(s)$ (for $N$ even) or $G(s)=G_o(s)$ (for $N$
odd). If $s$ is not the lexicographically smallest element in $G(s)$,
remove it from the list. The remaining elements are a unique
representation of the symmetry classes.

\section*{References}

\bibliographystyle{unsrt} 
\bibliography{labs,mertens,math,complexity}

\end{document}